\documentclass[aps,prb,twocolumn,showpacs,superscriptaddress,floatfix]{revtex4}
\usepackage[latin1]{inputenc}
\usepackage{amsfonts}
\usepackage[intlimits]{amsmath}
\usepackage{bbm}
\usepackage{stmaryrd}
\usepackage{graphicx}
\usepackage[bookmarks,pdftitle={Matrix product state approach for a two-lead,
  multi-level Anderson impurity model}]{hyperref}
\usepackage{xcolor}

\newcommand{\id}{\mathbbm{1}}
\DeclareMathOperator{\tr}{tr}
\def\bra#1{\ensuremath{\langle{#1}\vert}}
\def\ket#1{\ensuremath{\vert{#1}\rangle}}
\def\bracket#1#2{\ensuremath{\langle{#1}\mkern1.2mu\vert\mkern1.2mu{#2}\rangle}}
\def\expect#1{\ensuremath{\langle{#1}\rangle}}
\def\figref#1{\figurename~\ref{#1}}

\renewcommand{\figurename}{Fig.}

\begin{document}

\title{Matrix product state approach for a two-lead, multi-level  Anderson impurity model}

\author{Andreas \surname{Holzner}}
\affiliation{Physics Department, Arnold Sommerfeld Center for Theoretical Physics, and Center for NanoScience,
  Ludwig-Maximilians-Universit\"at M\"unchen, D-80333 M\"unchen, Germany}
\affiliation{Institute for Theoretical Physics C, RWTH Aachen University, D-52056 Aachen, Germany}
\author{Andreas \surname{Weichselbaum}}
\affiliation{Physics Department, Arnold Sommerfeld Center for Theoretical Physics, and Center for NanoScience,
  Ludwig-Maximilians-Universit\"at M\"unchen, D-80333 M\"unchen, Germany}
\author{Jan \surname{von Delft}}
\affiliation{Physics Department, Arnold Sommerfeld Center for Theoretical Physics, and Center for NanoScience,
  Ludwig-Maximilians-Universit\"at M\"unchen, D-80333 M\"unchen, Germany}

\date[Date: ]{April 3, 2008}

\begin{abstract}
  We exploit the common mathematical structure of the numerical renormalization group and the density matrix
  renormalization group, namely, matrix product states, to implement an efficient numerical treatment of a two-lead,
  multi-level Anderson impurity model. By adopting a star-like geometry, where each species (spin and lead) of
  conduction electrons is described by its own Wilson chain, instead of using a single Wilson chain for all species
  together, we achieve a very significant reduction in the numerical resources required to obtain reliable results. We
  illustrate the power of this approach by calculating ground state properties of a four-level quantum dot coupled
  to two  leads. The success of this proof-of-principle calculation suggests that the star geometry constitutes a
  promising strategy for future calculations the ground state properties of multi-band, multi-level quantum impurity
  models. Moreover, we show that it is possible to find an ``optimal'' chain basis, obtained via a unitary
  transformation (acting only on the index distinguishing different Wilson chains), in which degrees of freedom on
  different Wilson chains become effectively decoupled from each other further out on the Wilson chains. This basis
  turns out to also diagonalize the model's chain-to-chain scattering matrix. We demonstrate this for a spinless
  two-lead model, presenting DMRG-results for the mutual information between two sites located far apart on different
  Wilson chains, and NRG results with respect to the scattering matrix.
\end{abstract}

\pacs{78.20.Bh, 
  02.70.-c, 
  72.15.Qm, 
  75.20.Hr 
}

\maketitle

\section{Introduction}
\label{sec:intro}

A very successful method for solving quantum impurity models is Wilson's numerical renormalization group (NRG)
\citep{krishna-murty:nrg_siam,wilson:nrg,BullaCostiPruschke2008}. Recently, it has been pointed out
\citep{verstraete:vmps} that the approximate eigenstates of the Hamiltonian produced by NRG have the structure of matrix
product states (MPSs).\citep{rommer:thd-dmrg-mps} This observation established a structural relation between NRG and the
density matrix renormalization group (DMRG) \citep{Schollwock2005,white:dmrg2,white:dmrg1} because the states produced
by the latter likewise have the form of
MPS.\citep{dukelsky1998,fannes1992,VerstraeteMartin-DelgadoCirac2004,verstraete:dmrg-pbc,TakasakiHikiharaNishino1999}

This structural relation between NRG and DMRG has opened up very interesting perspectives for combining  advantageous
features of both methods. In particular, the fact that DMRG, in essence, is a method for variationally optimizing
MPSs\citep{dukelsky1998,TakasakiHikiharaNishino1999,verstraete:dmrg-pbc} can be used to devise a corresponding
variational treatment of quantum impurity models.\citep{verstraete:vmps,SaberiWeichselbaumDelft2008} This has the
advantage that MPSs with much richer more complex structures can be adopted than those produced by standard NRG,
entailing a much more efficient use of numerical resources. Concretely, the dimension $D$ of the matrices from which the
MPS is constructed can be reduced very significantly, typically by several orders of magnitude. As a result, it becomes
feasible to study complex quantum impurity problems that would be very challenging for standard NRG.

\begin{figure}[ht]
  \centering
  \includegraphics[width=0.8\linewidth=1]{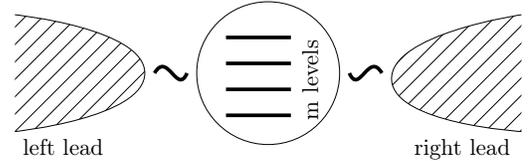}
  \caption{Quantum dot coupled to two leads.}
  \label{fig:quantum-dot}
\end{figure}
In this paper, we illustrate this idea by calculating ground state properties of a multi-level quantum dot coupled to
two spinful leads. Standard NRG treats the latter as a single quantum chain with $2^4$ states per site (to account for two
spin and two lead degrees of freedom), for which one typically needs $D \gtrsim 4000$ to achieve satisfactory results. In
contrast to the latter ``single-chain geometry,'' we adopt here a MPS with a ``star geometry,'' involving four separate
chains, each with only two states per site, and variationally optimize one chain after the other. This enables us to
obtain good results using matrices with $D$ ranging between 16 and 36. This reduction in numerical memory resources
relative to standard NRG illustrates the increased numerical efficiency alluded to above. Furthermore, we show that a
numerically optimal basis, involving rotated Wilson chains, can be found by requiring that this new representation
minimizes the mutual information between different chains. This optimal basis has an instructive physical
interpretation: it is the basis in which the chain-to-chain scattering matrix is diagonal.

This paper is structured as follows. In Sec.~\ref{sec:matrix-product-state-ansatz} we briefly review why standard NRG
produces MPSs with a single-chain geometry and advocate the adoption of MPSs with an alternative star geometry. In
Sec.~\ref{sec:vari-optim-scheme}, we describe how a star-MPS representation of the ground state can be determined by
variationally minimizing its energy. In Sec.~\ref{sec:results-occ} we present proof-of-principle calculations of some
ground-state properties and comparisons thereof to NRG results. Finally, Sec.~\ref{sec:rotate-leads} illustrates how
a numerically optimal basis for the chains can be obtained by effectively minimizing the mutual information between two
sites of different chains.

\section{Matrix product state Ansatz}
\label{sec:matrix-product-state-ansatz}

\subsection{Model}
\label{sec:model}

We study a multi-level, two-lead Anderson impurity model described by the following Hamiltonian:
\begin{equation}
  \label{eq:h-total}
  H = H_{\rm dot} + H_{\rm int} + H_{\rm leads} + H_{\rm coupling},
\end{equation}
where $H_{\rm dot}$ describes the eigenenergies of the $m$ dot levels
\begin{equation}
  \label{eq:h-dot}
  H_{\rm dot} = \sum_{i=1}^{m}\sum_{s=\shortuparrow,\shortdownarrow} \epsilon_{is} d^\dagger_{is} d^{\phantom{\dagger}}_{is},
\end{equation}
$H_{\rm int}$ is the Coulomb interaction on the dot
\begin{equation}
  \label{eq:h-int}
  H_{\rm int} = \frac{U}{2} \sum_{(i,s)\neq(j,s')} d^\dagger_{is} d^{\phantom{\dagger}}_{is} d^\dagger_{js'}
  d^{\phantom{\dagger}}_{js'},
\end{equation}
$H_{\rm leads}$ is the free lead Hamiltonian for $N_{\rm l}$ leads ($\alpha = 1, \ldots, N_{\rm l}$)
\begin{equation}
  \label{eq:h-lead}
  H_{\rm leads} = \sum_{\vec{k} \alpha s} \epsilon_{\vec{k}} c^{\dagger}_{\vec{k}\alpha s}
  c^{\phantom{\dagger}}_{\vec{k}\alpha s},
\end{equation}
and $H_{\rm coupling}$ is the coupling between the dot levels and the leads
\begin{equation}
  \label{eq:h-coupling}
  H_{\rm coupling} = \sum_{i\vec{k}\alpha s} V_{i\alpha} \left(d^{\dagger}_{is} c^{\phantom{\dagger}}_{\vec{k}\alpha s}
    + c^{\dagger}_{\vec{k}\alpha s} d^{\phantom{\dagger}}_{is} \right).
\end{equation}
At a late stage of this work we became aware of work of Kashcheyevs et~al.\citep{Kashcheyevs2007} suggesting to perform
a singular value decomposition on $H_{\rm coupling}$ which has the merit of decoupling some levels from some
leads. Applying this idea to our system should also give some improvement in numerical efficiency. In general, however,
all the levels will remain to be coupled to all leads. As we will show later, a more general scheme than just a singular
value decomposition is capable of generating a new basis for the leads that will minimize the coupling of the leads
amongst themselves.

Following Wilson, \citep{krishna-murty:nrg_siam} we adopt a logarithmic discretization of the conduction bands 
and tridiagonalize 
$H_{\rm leads} + H_{\rm coupling}$. As a result, the dot, represented by the ``dot site,'' is coupled to the first sites
of $2 N_{\rm l}$ separate ``Wilson'' chains, labeled by $(\alpha, s)$
\begin{gather}
  \label{eq:h-coupling-nrg}
  H_{\rm coupling} = W \sum_{i\alpha s} \sqrt{\frac{2\Gamma_{i\alpha}}{\pi W}} \left(f^{\dagger}_{0\alpha s}
    d^{\phantom{\dagger}}_{is} + d^{\dagger}_{is} f^{\phantom{\dagger}}_{0\alpha s}\right)\\
  \begin{split}
    H_{\rm leads} &= W \sum_{\alpha s} \frac{1}{2} (1 + \Lambda^{-1})\\[-1ex]
    &\times\sum_{n=0}^{L-1} \Lambda^{-\frac{n}{2}} \xi_{n}
    \left( f^{\dagger}_{n\alpha s\phantom{()}} \negmedspace f^{\phantom{\dagger}}_{(n+1)\alpha s} + \text{H.c.} \right).
  \end{split}\label{eq:h-leads-nrg}
\end{gather}
Here $\xi_{n} = (1 - \Lambda^{-n-1})(1 - \Lambda^{-2n-1})^{-\frac{1}{2}}(1 - \Lambda^{-2n -3})^{-\frac{1}{2}}$ are
coefficients of order 1, $\Gamma_{i\alpha} = \pi \rho V^{2}_{i\alpha}$ the hybridization, $\rho$ is the density of
states, and $2W$ is the bandwidth of the conduction bands of the leads centered at the Fermi edge. We set the NRG
discretization parameter $\Lambda = 2$ throughout this paper.  The length $L$ of the Wilson chain determines the energy
resolution with which the lowest-lying eigenstates of the chain are resolved. We typically choose $L=60$.

A standard NRG treatment of this model would combine all four Wilson chains into a single
one, whose sites are labeled by a single site index $k= 0, \dots, L$ [see \figref{fig:different-geometries}(a)]. Each
site would represent a $2^{2 N_{\rm l}}$-dimensional local 
state space, consisting of the set of states $\{ \ket{\sigma_{k}} \}$, where the state label $\sigma_k$ takes on
$2^{2 N_{\rm l}}$ different values.  Then one proceeds to diagonalize the Hamiltonian iteratively, as follows: suppose
a short Wilson chain up to and including site $k-1$ has been diagonalized exactly, yielding a set of eigenstates 
$\ket{i_{k}} \in {\rm span} \left\{ \{\ket{\sigma_{1}}\} \otimes \{\ket{\sigma_{2}}\} \otimes  \ldots \otimes
  \{\ket{\sigma_{k-1}}\} \right\}$.  Then one adds the next site, $k$, to the chain, thereby enlarging the Hilbert space
by a factor of $2^{2 N_{\rm l}}$, diagonalizes the Hamiltonian in this enlarged space, and truncates by discarding 
all but the lowest $D$  eigenstates of the Hamiltonian. The latter can in general be written
\begin{figure}[ht]
  \centering
  \includegraphics[width=0.9\linewidth]{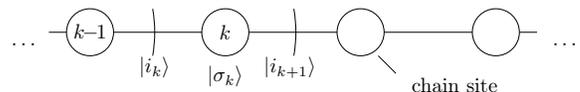}
  \caption{Iterative generation of matrix product states for a chain.}
  \label{fig:basis-chain}
\end{figure}
as linear combinations of the following form (illustrated in \figref{fig:basis-chain}):
\begin{equation}
  \label{eq:nrg-combined-basis}
  \ket{i_{k+1}} = \sum_{i_{k}, \sigma_{k}} A^{[\sigma_{k}]}_{i_{k},i_{k+1}} \ket{i_{k}}\ket{\sigma_{k}}.
\end{equation}
Iterating this procedure up to and including site $L$ produces eigenstates of the form
\begin{equation}
  \label{eq:nrg-chain-mps}
  \ket{i_{L+1}} = A^{[\sigma_{k}]}_{i_{k},i_{k+1}} \ldots
  A^{[\sigma_{L}]}_{i_{L},i_{L+1}} \ket{i_{k}}\ket{\sigma_{k}} \ldots \ket{\sigma_{L}},
\end{equation}
where sums over repeated indices are implied. Since such states are completely characterized by sums over products of
matrices, they have come to be known as matrix product states. The form of these MPS produced by NRG is analogous to the
state for a chain as shown in \figref{fig:mps-graph}.

\subsection{Star geometry}
\label{sec:mps}

One limiting factor for the accuracy of the NRG approach is that a certain amount
of information is lost at each iteration step due to truncation. 
In general, for a system with $N_{\rm l}$ bands (in the two-lead case which we will investigate below,  $N_{\rm l} = 2$),
the dimension of the effective Hilbert space is enlarged from $D$ to $D 2^{2 N_{\rm l}}$ upon adding a new site to the Wilson
chain.  Thus, the larger $N_{\rm l}$, the more information is lost during the subsequent truncation of
the Hilbert space back to dimension $D$, and the less accurate the NRG treatment is expected to be. 

\begin{figure}[ht]
  \centering
  \includegraphics[width=0.8\linewidth]{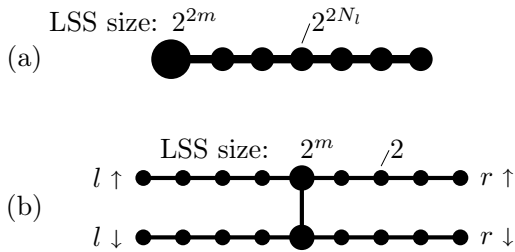}
  \caption{(a) Single chain geometry: a single Wilson chain of local dimension $2^{2N_{\rm l}}$ coupled to one dot site of
    local dimension $2^{2m}$. (b) Star geometry: $2N_{\rm l}$ Wilson chains (here $N_{\rm l} = 2$ and $\alpha = l,r$) ,
    each with local dimension 2, coupled to two dot sites of local dimension $2^{m}$.}
  \label{fig:different-geometries}
\end{figure}
The main goal of the present paper is to illustrate that a very significant improvement of efficiency can be obtained
as follows: instead of combining all $2 N_{\rm l}$ chains into a single
Wilson chain of local dimension $2^{2 N_{\rm l}}$ (``single-chain geometry''),
we shall treat them as separate  chains, each with local dimension 2 and each coupled to the 
same set of dot levels [``star geometry,'' see \figref{fig:different-geometries}(b)]. Although the total number of sites thereby 
increases from ${\cal O} (L)$ to ${\cal O}(N_{\rm l} L)$, the dimension of the local state 
space \emph{per site} is reduced from $2^{2 N_{\rm l}}$ to 2. We find that, due to the latter fact,
the dimension $D$ of the constituent matrices in the star-MPS can be chosen
to be significantly smaller than in the chain-MPS.

The change from single-chain to star geometry, 
however, necessitates  a change in truncation strategy for the following
reason: in contrast to the single-chain geometry,
where each site represents a definite energy scale, in the star geometry a given scale 
is represented by a set of $2 N_{\rm l}$ sites, one on each of the star's chains, i.e., at
locations that are widely ``separated'' from each other on the star. 
Therefore, a truncation scheme based on energy scale separation, 
such as that used by standard NRG, can no longer be applied. 
Instead, we shall simply minimize \cite{verstraete:vmps}  the expectation value of the
Hamiltonian within the space of all MPSs with the same star structure.
This can be done efficiently by  optimizing the matrices in the star-MPS 
one site at a time, and sweeping through all sites until convergence. 

\begin{figure}[ht]
  \centering
  \includegraphics[width=0.9\linewidth]{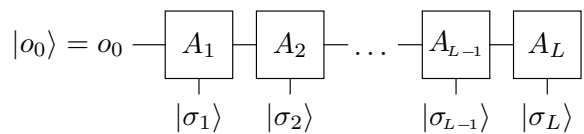}
  \caption{Graphical representation of Eq.~\eqref{eq:chain-mps}.}
  \label{fig:mps-graph}
\end{figure}

To be explicit, we construct our star-MPS for the two-lead system as follows.
In total $4 = 2 N_{\rm l}$ ($N_{\rm l} = 2$) Wilson chains are connected to the dot. Each of these chains is very
similar to the NRG MPS from above, except that the local state space (LSS) is only of dimension 2.
To simplify the notation
we drop the labels $\alpha$ and $s$ whenever possible and incorporate them into the site index $k$, which from now on
will be taken to uniquely determine a site in the whole star structure. $\sigma_{k}$ still labels the LSS at site $k$.
With this every Wilson chain can be represented as (see \figref{fig:mps-graph})
\begin{subequations}
  \begin{align}
    \label{eq:chain-mps}
    \ket{o_{0}} &= A^{[\sigma_{1}]}_{o_{0} o_{1}} A^{[\sigma_{2}]}_{o_{1} o_{2}} \ldots A^{[\sigma_{L}]}_{o_{L-1}}
    \ket{\sigma_{1}} \ket{\sigma_{2}} \ldots \ket{\sigma_{L}},\\
    \label{eq:chain-mps-shorthand}
    &= \left( \prod_{k=1}^{L} A^{[\sigma_{k}]} \right) \ket{\vec{\sigma}}
  \end{align}
\end{subequations}
where $\ket{\vec{\sigma}} = \ket{\sigma_{1}}\ket{\sigma_{2}} \ldots \ket{\sigma_{L}}$. Here the label $o$ stands for
``outer,'' for reasons that will become clear below.
We introduce an intuitive graphical representation for these MPS. Every $A$ will be represented by a box and every index
of $A$ is depicted by a line attached to the box. For matrix products or other index summations the corresponding lines
are connected. Using this representation, a single chain can be depicted as in \figref{fig:mps-graph}.

\begin{figure*}[ht]
  \includegraphics[width=0.9\textwidth]{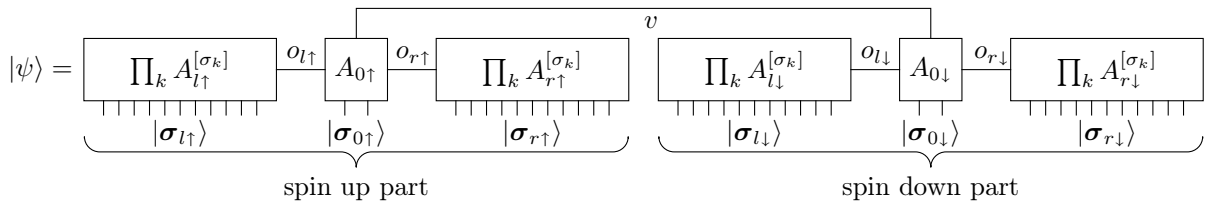}
  \caption{MPS representation for a quantum dot coupled to two spinful leads. The lead chains are combined to big
    boxes for clarity. The indices of the dot matrices are labeled explicitly.}
  \label{fig:mps-whole-system}
\end{figure*}

The fact that the Hamiltonian does not contain terms that flip spin up to down or vice versa
suggests representing the dot state space by two separate sites, representing all dot states having spin up or down,
respectively [see \figref{fig:different-geometries}(b)]. Correspondingly, we also introduce two types of dot matrices,
$A^{[\sigma_{0\shortuparrow}]}$ and $A^{[\sigma_{0\shortdownarrow}]}$, which carry an extra index $v$ that is being
summed over to link the spin up and down subsystems. So we arrive at the starlike structure of
\figref{fig:mps-whole-system} with two linked dot matrices (one for each spin) and two leads (left and right) attached
to each:
\begin{equation}
  \label{eq:mps-whole-system-explicit}
  \begin{split}
    \ket{\psi} = &\left( \prod_{k_{l\shortuparrow}} A^{[\sigma_{k_{l\shortuparrow}}]} \!\right)_{\!\negthickspace
      o_{l\shortuparrow}} \negthickspace\negthickspace A^{[\vec{\sigma}_{0\shortuparrow}]}_{o_{l\shortuparrow} o_{r\shortuparrow} v}
    \!\left( \prod_{k_{r\shortuparrow}} A^{[\sigma_{k_{r\shortuparrow}}]} \!\right)_{\!\negthickspace o_{r\shortuparrow}}\\
    &\left( \prod_{k_{l\shortdownarrow}} A^{[\sigma_{k_{l\shortdownarrow}}]} \!\right)_{\!\negthickspace o_{l\shortdownarrow}}
    \negthickspace\negthickspace A^{[\vec{\sigma}_{0\shortdownarrow}]}_{o_{l\shortdownarrow} o_{r\shortdownarrow} v} \!\left(
      \prod_{k_{r\shortdownarrow}} A^{[\sigma_{k_{r\shortdownarrow}}]} \!\right)_{\!\negthickspace o_{r\shortdownarrow}}\\
    &\;\ket{\vec{\sigma}_{l\shortuparrow}} \ket{\vec{\sigma}_{0\shortuparrow}} \ket{\vec{\sigma}_{r\shortuparrow}}
    \ket{\vec{\sigma}_{l\shortdownarrow}} \ket{\vec{\sigma}_{0\shortdownarrow}} \ket{\vec{\sigma}_{r\shortdownarrow}}.
  \end{split}
\end{equation}
This starlike structure basically consists of two y-junctions, as discussed by Guo and White,\citep{Guo2006}
next to each other.

Hiding the explicit structure [Eq.~\eqref{eq:mps-whole-system-explicit}] of the MPS as illustrated in
\figref{fig:mps-whole-system}, we can write a state symbolically as
\begin{equation}
  \label{eq:psi-global}
  \ket{\psi} = \left( \prod_{k} A^{[\sigma_{k}]} \right) \ket{\vec{\sigma}}.
\end{equation}
We call Eq.~\eqref{eq:psi-global} the global representation of $\ket{\psi}$. 

An important point to note is that this system is still effectively one dimensional, in the sense that if we cut out a
given site, the system breaks apart into two (or three in case of a dot site) disjoint parts. We shall call the one
containing the dot sites the ``inner'' part, the other one the outer part. As a consequence, it is possible to also
give a ``local'' description of $\ket{\psi}$ of the form
\begin{equation}
  \label{eq:psi-local}
  \ket{\psi} = A^{[\sigma_{k}]}_{i_{k},o_{k}} \ket{i_{k}}\ket{\sigma_{k}}\ket{o_{k}},
\end{equation}
where $\{\ket{\sigma_{k}}\}$ represents
the LSS of the chosen site, $\{\ket{i_{k}}\}$ is an orthonormal set of states representing the inner state space (ISS),
namely, the inner part of the star with respect to the chosen site $k$, and $\{\ket{o_{k}}\}$ is an orthonormal set of
states representing  the outer state space (OSS), namely, the outer part of the star.

\section{Variational site optimization scheme}
\label{sec:vari-optim-scheme}

We will use the MPS of \figref{fig:mps-whole-system} as an ansatz for the ground state of our
system. In order to find the ground state we need to calculate the MPS $\ket{\psi}$ that minimizes the energy $E =
\bra{\psi} H \ket{\psi}$ with the constraint of keeping the norm of $\ket{\psi}$ constant.\citep{verstraete:vmps} Using
$\lambda$ as Langrange multiplier ensuring normalization we arrive at the following minimization problem:
\begin{equation}
  \label{eq:min-gs}
  \min_{\ket{\psi}} \left( \bra{\psi} H \ket{\psi} - \lambda \bracket{\psi}{\psi} \right).
\end{equation}
The key idea of the variational MPS optimization is to optimize every single $A$-matrix of $\ket{\psi}$ separately until
the ground-state energy has converged. Therefore we insert the local MPS description from Eq.~\eqref{eq:psi-local} into
Eq.~\eqref{eq:min-gs} and obtain
\begin{equation}
  \label{eq:min-A}
  \min_{A_{k}} \left( A^{[\sigma'_{k}]*}_{i' o'} H_{(i'\sigma'_{k} o'), (i \sigma_{k} o)} A^{[\sigma_{k}]}_{i o} -
    \lambda A^{[\sigma_{k}]*}_{i o} A^{[\sigma_{k}]}_{i o} \right),
\end{equation}
where $H_{(i' o' \sigma'_{k}), (i o \sigma_{k})}$ are the Hamilton matrix elements in the current effective bases
\begin{equation}
  \label{eq:ham-mat-elem}
  H_{(i' \sigma'_{k} o'), (i \sigma_{k} o)} = \bra{o'}\bra{\sigma'_{k}}\bra{i'} H \ket{i}\ket{\sigma_{k}}\ket{o}.
\end{equation}
By setting the derivative of Eq.~\eqref{eq:min-A} with respect to the matrix elements of $A_{k}^{*}$ to zero and replacing
$\lambda$ by $E_{o}$, we obtain the following eigenvalue equation for $A_{k}$:
\begin{equation}
  \label{eq:opt-eigv}
  H_{(i' \sigma'_{k} o'), (i \sigma_{k} o)} A^{[\sigma_{k}]}_{i o} = E_{0} A^{[\sigma'_{k}]}_{i' o'}.
\end{equation}
The eigenvector with the smallest eigenvalue is the solution to our minimization problem. So after having solved this
eigenvalue problem for the current site $k$ we replace $A_{k}$ with the newly found eigenvector and move on to the next
site in order to optimize that $A_{k'}$. We repeat the whole process (sweeping) until the ground-state energy has
converged (see below).

By following this procedure we succeed to divide a very high dimensional minimization problem into manageable smaller
units. For general problems this can be a very bad approach as one can get stuck in a local minimum during the
optimization. However, it has proven to work reliably when the site-site coupling varies smoothly and monotonously. In
our case the Hamiltonian has only nearest-neighbor interactions and there are no long-range correlations in the
system. As a result, the system reliably converges without getting stuck in local minima.

\subsection{Updating the $A$ matrices and changing the effective basis states}
\label{sec:effective-basis}

When updating $A$ matrices during sweeping, one must ensure that two conditions are satisfied. First,
whenever we use the local description of Eq.~\eqref{eq:psi-local}, we rely on the basis states  being orthonormal:
$\bracket{o_{k}}{o'_{k}} = \delta_{o_{k},o'_{k}}$. This condition translates to
\begin{equation}
  \label{eq:mat-outer-basis}
  \sum_{\sigma_{k'}} A^{[\sigma_{k'}]} A^{[\sigma_{k'}]\dagger} = \id \qquad\text{for } k' > k,
\end{equation}
for all outer matrices with respect to site $k$. We will focus here on the
OSS basis, everything works completely analogously for the ISS basis.

Second, we also want to create an effective basis that spans a DMRG optimal Hilbert space, i.\,e.,\ the states we
keep for an effective basis are to be the ones having the largest weights in the density matrix of the current
state (as described below).

For definiteness, we consider an inward sweep and focus on how to move the ``current site'' from $k$ to $k-1$. We
assume that a new set of $A$ matrices for site $k$ has been obtained by energy minimization. The question is how to
ensure that both above mentioned conditions are satisfied. As all the inner $A$ matrices of site $k - 1$ have not
changed since we optimized site $k-1$ the last time when  moving outwards, we only need to create a new effective OSS
basis $\ket{o_{k-1}}$ for site $k-1$.

Starting from the density matrix in the local description of site $k$,
\begin{equation}
  \label{eq:dm-local}
  \rho^{(k)} = \ket{\psi}\!\bra{\psi} = A_{i^{\phantom{*}} o}^{[\sigma_{k}]} A_{i' o'}^{[\sigma'_{k}] *}
  \ket{i}\!\bra{i'} \ket{\sigma_{k}}\!\bra{\sigma'_{k}} \ket{o}\!\bra{o'},
\end{equation}
suppose one traces out the inner part of this system to obtain reduced density matrix of the outer part and site $k$,
\begin{equation}
  \label{eq:dm-red}
    \rho^{(k)}_{\rm red} = \tr_{i}\rho^{(k)} = A_{i^{\phantom{*}} o}^{[\sigma_{k}]} A_{i o'}^{[\sigma'_{k}] *}
    \ket{\sigma_{k}}\!\bra{\sigma'_{k}} \ket{o}\!\bra{o'}
\end{equation}
which corresponds precisely to the outer part with respect to site $k-1$.

Now employ the singular value decomposition (SVD) $A = U S V^{\dagger}$ which exists for every
rectangular matrix $A$. $S$ is a diagonal matrix containing the singular values ordered by magnitude; $U$ and
$V^{\dagger}$ are column and row unitary matrices, respectively, and obey $U^{\dagger}U = V^{\dagger}V = \id$.
Combine $\ket{\sigma_{k}}$ and $\ket{o_{k}}$ to $\ket{l_{k}} = \ket{\sigma_{k}}\ket{o_{k}}$ and insert the SVD
for $A_{i l} = U_{i m} S_{m j} (V^{\dagger})_{j l}$
\begin{equation}
  \label{eq:dm-svd}
  \begin{split}
    \rho^{(k)}_{\rm red} &= A_{i l}^{\phantom{*}} A_{i l'}^{*} \ket{l_{k}}\!\bra{l'_{k}} = V_{j' l'} S_{l' m} S_{m l} V^{\dagger}_{j l}
    \ket{l_{k}}\!\bra{l'_{k}}\\
    &= \sum_{j} \rho^{(k)}_{j} \ket{j_{k}}\!\bra{j_{k}}.
  \end{split}
\end{equation}

The second line follows since $S^{2}$ is diagonal, and we wrote $\rho^{(k)}_{j} = S^{2}_{j j}$ and $\ket{j_{k}} =
V^{\dagger}_{j l} \ket{l_{k}}$.
We see that the SVD automatically diagonalizes the reduced density matrix with the states ordered according to
their weight.

\begin{figure}[ht]
  \centering
  \includegraphics[width=\linewidth]{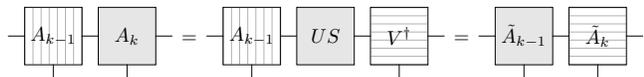}
  \caption{Procedure for moving the actual site from $k$ to $k-1$. The matrices that are not orthonormalized in any
    direction are printed with gray background. The gray lines within the boxes indicate whether the row or column
    vectors are orthonormal (with the local level associated with row or column, respectively).}
  \label{fig:mps-move-site}
\end{figure}

So all we actually have to do for moving the actual site from $k$ to $k-1$ is to calculate the SVD of the newly
optimized $A_{k} = U S V^{\dagger}$. We then replace $A_{k} \rightarrow \tilde{A}_{k} = V^{\dagger}$ and $A_{k-1}
\rightarrow \tilde{A}_{k-1} = A_{k-1} U S$ as illustrated by \figref{fig:mps-move-site}.
By doing so we do not change the total state, since the product
\begin{equation}
  \label{eq:update-const-product}
A_{k-1} A_{k} = \tilde{A}_{k-1} \tilde{A}_{k}
\end{equation}
remains unchanged. Thus we create an effective orthonormal OSS basis,
\begin{equation}
  \label{eq:update-new-basis}
  \ket{o_{k-1}} = \tilde{A}^{[\sigma_{k}]}_{o_{k-1} o_{k}} \ket{\sigma_{k}}\ket{o_{k}}
\end{equation}
which at the same time is DMRG optimal.

The so-called site optimization procedure outlined above, where we optimize the $A$ matrices directly, is equivalent to
one-site finite-size DMRG.

The relation between the singular values and the weights of the reduced density matrix can be used to optimize our
choice for the dimensions of the respective effective Hilbert spaces: instead of using the same dimensions for
\emph{all} $A$ matrices in the system, which turns out to be inefficient for inhomogeneous ones like ours, we adopt as
truncation criterion the demand that the minimum value of $S^{2}$ at a given site is to be smaller than some
threshold $w_{\rm min}$ (in our case typically taken as $10^{-6}$). After calculating the singular values, we choose the
matrix dimensions $D_{k}$ at the corresponding bond $k$ (between site $k$ and its neighbor in the direction of the dot)
according to the following recipe. We choose $D_{k}$ large enough to ensure that the minimal singular value $s_{\rm
  min}(k)$ fulfills $s_{\rm min}^{2}(k) < w_{\rm min}$, but subject to this constraint choose $D_{k}$ to be as small as
possible, in order to minimize computational resources.

\begin{figure}[ht]
  \centering
  \includegraphics[width=\linewidth]{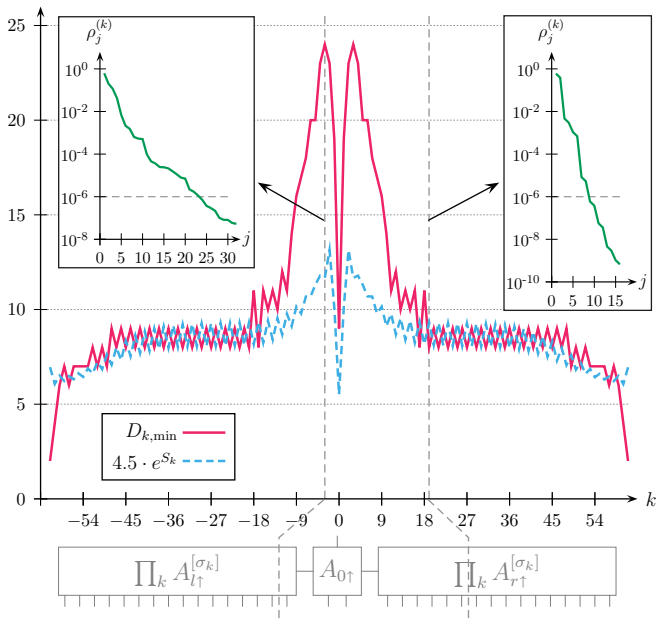}
  \caption{(Color online) The solid line shows the dimension $D_{k}$ needed at bond $k$ of the spin up chain to satisfy
    $w_{\rm min} = 10^{-6}$ for the reduced density matrix at each bond (negative $k$ correspond to the left chain). The
    dashed line displays the exponentiated bond entropy $e^{S_{k}}$ multiplied by $4.5$ to visually match the $D_{k, \rm
      min}$ curve for large $k$. Here $k = 0$ corresponds to the ``vertical'' bond between the two spin subsystems. The
    two insets show spectra of reduced density matrices at different bonds $k$ indicated by the vertical dashed lines of
    the main plot. The data shown in this figure has been obtained from the ground state of the four-level model shown
    in \figref{fig:4l-occ} with $\epsilon = -1.7 U$. In general, the maximum dimension needed depends strongly on the
    model parameters.}
  \label{fig:bond-dim}
\end{figure}

It is instructive to also explore the relation between $D_{k}$ and the bond entropy $S_{k}$ of site $k$, which can be
computed from the reduced density matrix $\rho_{\rm red}^{(k)}$ at site $k$ according to
\begin{equation}
  \label{eq:bond-entropy}
  S_{k} = -\tr\left(\rho_{\rm red}^{(k)}\ln\rho_{\rm red}^{(k)}\right).
\end{equation}
The entropy $S_{k}$ is a measure for the entanglement between the traced out part of the system and the part kept in
the description of $\rho_{\rm red}^{(k)}$. Thus, large $S_{k}$ implies large $D_{k}$, which turns out to be roughly
proportional to $e^{S_{k}}$. The dimensions $D_{k}$ resulting from the above criterion for the singular values $s_{\rm
  min}(k)$ together with the exponentiated bond entropy $e^{S_{k}}$ associated with the reduced density matrix at bond
$k$ are shown in \figref{fig:bond-dim}. This figure shows, first, that a larger dimension is required near the dot,
and second, that $e^{S_{k}}$ (times a constant) is a rather good indicator of the required dimension $D_{k}$. For the
limiting case of a reduced density matrix $\rho_{\rm red}^{(k)}$ with uniform weights $\rho_{j}(k) = \frac{1}{D_{k}}
\forall j\in\left[1,D_{k}\right]$, the exponentiated bond entropy then gives $e^{S_{k}} = D_{k}$. Thus, $D_{k}$ is a
upper bound to $e^{S_{k}}$.\cite{NielsonChuang2000} The dip at $k = 0$ for the bond between the two spin subsystems
(dimension $D_{v}$) is due to the fact that there is only a density-density interaction along this bond but no particle
exchange. For our system we found that it is sufficient to have dimensions of 36 or less near the dot.

\subsection{Sweeping sequence}
\label{sec:sweeping-seq}

\begin{figure}[ht]
  \centering
  \includegraphics[width=0.9\linewidth]{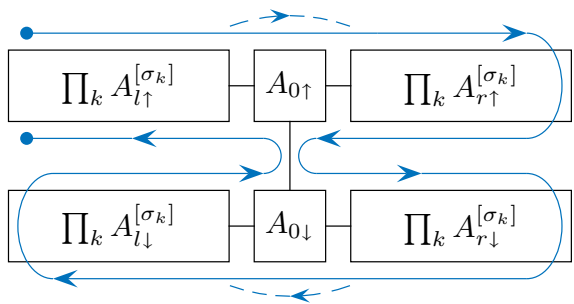}
  \caption{(Color online) Sweeping sequence. For clarity we place the spin up and spin down parts on top of each other
    to emphasize the star-like structure. The solid blue line depicts the standard sweeping sequence.}
  \label{fig:sweeping-seq}
\end{figure}

In principle the order in which we optimize the single matrices during a sweep is not important. However, it is both
convenient and more efficient to move only to a neighboring site (and not further) for the next optimization step. In
this way we need to change the actual site only by one in order to get the desired new local description. Having our MPS
ansatz structure in mind, this requirement immediately suggests a particular order of sweeping, shown in
\figref{fig:sweeping-seq}. Starting from the far end of any chain we move in toward the dot matrix and then out again
along another chain. We repeat this until we have covered the whole system.
Sweeping that way (solid blue line in \figref{fig:sweeping-seq}) we optimize the two dot matrices three times but all
the other sites only twice. If one wants to optimize all sites twice during a sweep one can once skip the optimization
step at the dot sites as indicated by the dashed blue line.

As the dot matrices are by far the biggest in the system, optimizing them takes much longer than optimizing any of the
chain matrices. Thus by skipping the dot optimization step once, we can reduce the computational time needed for a
single sweep. However, since the dot optimization step also has the biggest effect for improving our MPS ansatz,
skipping its optimization once has to be compensated by performing more sweeps to achieve as good convergence of the
ground state as in the case where we perform three optimizations at the dot matrices. We compared both approaches for
our model and found no significant differences in the overall performance.

We stop the sweeping when the MPS has converged. To probe the convergence we compare the MPSs before and after sweep
$N$, $\ket{\psi_{N-1}}$, and $\ket{\psi_{N}}$. If the change in overlap,
\begin{equation}
  \label{eq:sweep-crit}
  1 - | \bracket{\psi_{N-1}}{\psi_{N}} | \leq \epsilon,
\end{equation}
is smaller than a certain threshold, we stop the sweeping. We typically use $\epsilon = 10^{-3}$ and need 10 -- 15
sweeps. This depends crucially on the system parameters, though, and in some cases we need to perform up to 25 sweeps.

\subsection{Numerical costs}
\label{sec:numerical-costs}

The most computational effort is needed for solving the eigenvalue problem [Eq.~\eqref{eq:opt-eigv}] for the minimal
eigenvector. We use the Lanczos method for solving Eq.~\eqref{eq:opt-eigv}, which is an iterative method and requires the
calculation of $H \ket{\psi}$ in the local picture once for every iteration. As we cannot influence the number of
Lanczos iterations in our implementation, we will only investigate the costs of calculating $H \ket{\psi}$, which are
given by the costs of the matrix-matrix multiplication
$\sum_{i o \sigma_{k}} H_{(i'o'\sigma'_{k}),(i o \sigma_{k})} A_{i o}^{[\sigma_{k}]}$. The costs of a matrix-matrix
multiplication is given by the size of the outcome times the dimension of the index being summed over.
$H_{(i'o'\sigma'_{k}),(i o \sigma_{k})}$ splits up into a sum of different terms, such as
$(c^{\dagger}_{k})_{\sigma'_{k} \sigma_{k}} \otimes (c^{\phantom{*}}_{k+1})_{o' o}$, each consisting of a direct tensor
product of operators living in the ISS, OSS or LSS. Thus the product $H \ket{\psi}$ can be split up into smaller matrix
products. By looking at the structure of the Hamiltonian \eqref{eq:h-total}, one recognizes that there will be no terms
containing tensor products of operators from the ISS and OSS, since they would correspond to next-nearest-neighbor
terms, but tensor products with one operator from the LSS and the other one from the ISS or OSS. These terms lead to
multiplications over an index of length $Dd$, being the product of the dimensions of the ISS and LSS.
If the current site is the dot site, the size of the resulting matrix is $D^{2}D_{v}d^{m}$ and thus the costs for a
single multiplication $H\ket{\psi}$ at a dot site is given by
\begin{equation}
  \label{eq:cost-mult}
  C_{\rm dot} = \mathcal{O}(D^{2}D_{v}d^{m}Dd) = \mathcal{O}(D^{3}D_{v}d^{m+1}).
\end{equation}
In case of a chain site instead of a dot site exactly the same reasoning applies and because of the smaller matrix size
the costs reduce to $\mathcal{O}(D^{3}d^{2})$. From Eq.~\eqref{eq:cost-mult} we see that optimizing
the dot sites is the most expensive step in the optimization and scales particularly unfavorably when the number of dot
levels $m$ is increased.

\subsection{Bond optimization}
\label{sec:bond-optimization}

As an alternative to the site optimization scheme discussed above, we can begin to move the current site as in
\figref{fig:mps-move-site} to obtain $A_{k-1} (US) V^{\dagger}$, where $A_{k} = USV^{\dagger}$. At this step we can
represent the overall state as $\ket{\psi} = (US)_{i_{k}o_{k-1}} \ket{i_{k}}\ket{o_{k-1}}$. Now we perform the
optimization on $B = US$ in complete analogy to the site optimization and obtain a new $\tilde{B}$. Then $A_{k-1}$ is
replaced by $\tilde{A}_{k-1} = A_{k-1} \tilde{B}$ which results in a state with the actual site $k-1$. We call this
process ``bond optimization'' as the matrix we actually optimize is somehow located at the bond between two original
sites.

One can easily see that the costs for calculating $H_{(i'o'),(i o)} B_{i o}$ are $\mathcal{O}(D^{3})$ and thus
independent of the number of dot levels. Considering only the costs for a single sweep the bond optimization scheme will
be considerable faster than site optimization, which is especially expensive at the dot sites. This advantage, however,
is compromised to some extent by the slower convergence of the bond optimization due to the optimization taking place
within in a much smaller effective Hilbert space. This makes more sweeps necessary and also enforces a lower threshold
in Eq.~\eqref{eq:sweep-crit} as convergence criterion. It turned out to be very difficult to judge the convergence of the
bond optimization scheme based on Eq.~\eqref{eq:sweep-crit} especially if one starts from a state not too different from the
actual ground state because in such cases the convergence can be really slow and one might wrongly consider the state
already converged.

However, one might try to avoid unnecessary site optimizations at the beginning of the sweeping and use cheap bond
optimizations instead and switch after several sweeps to the site optimization scheme to make use of the better
convergence properties.

\section{Results for local occupations}
\label{sec:results-occ}

We used the approach described above to calculate the ground state and level occupancies of a spinful multi-level
quantum dot coupled to two leads. Throughout this part we fix the Coulomb interaction $U = 0.2 W$, $2W$ being the
bandwidth, and use the convention $W = 1$.

The results shown below demonstrate that it is possible to calculate local ground state quantities of a complex
quantum dot efficiently using this approach. Already with calculating the occupation of the dot levels it is possible
to investigate the stability diagram of small quantum dots.\citep{Gaudreau2006} Under certain conditions, local
occupancies can be related to phase shifts, which in turn can be used to calculate the conductance through a quantum
dot.\citep{Karrasch2007a}

\begin{figure}[t]
  \centering
  \includegraphics[width=\linewidth]{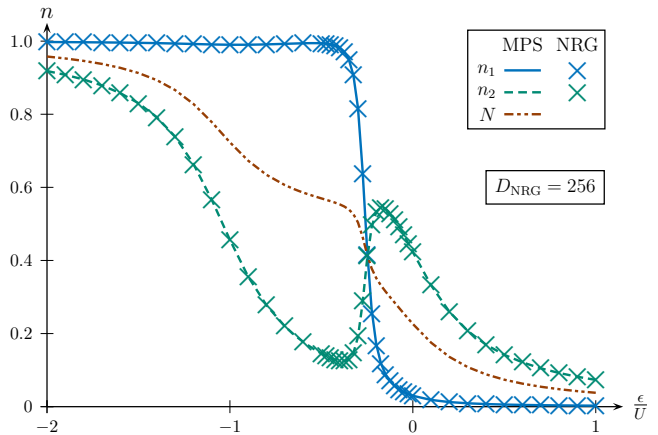}
  \caption{(Color online) Dot level occupation for a spinless two-level system, with $\epsilon_{1,2} = \epsilon \pm
    \Delta/2$, level spacing $\Delta = 0.1 U$ and couplings $\Gamma_{1l} = \Gamma_{1r} = 0.005 U$, $\Gamma_{2l} =
    \Gamma_{2r} = 30 \Gamma_{1l}$. This parameter set was used in Sindel {\em et~al.} (Ref.~\onlinecite{sindel2005}) $N
    = \frac{1}{2}(n_{1} + n_{2})$ is half the total dot occupation. Note that the sign in $\Gamma_{i\alpha}$ just serves
    as an indication of the sign of the related hopping matrix element $V_{i\alpha}$ in the Hamiltonian.}
  \label{fig:nrg-compare}
\end{figure}

First we consider the simpler case of a spinless two-level model with level positions $\epsilon_{1,2} =
\epsilon \pm \Delta/2$, coupled symmetrically to two leads. NRG works very reliable for this kind of impurity model. The
lower of the two levels is assumed to couple significantly stronger to the leads. We calculated the occupation, $n_{i} =
\expect{d_{i}^{\dagger}d_{i}^{\phantom{*}}}$, of both levels as a function of $\epsilon$, using both our MPS approach
and NRG. In \figref{fig:nrg-compare} we show the occupation of both levels as we sweep the gate potential by shifting
the levels from below towards the Fermi edge of the leads and then further above. At the beginning of this process
mainly the lower level starts to empty. This is due to the much bigger couplings $\Gamma_{2}$ of the lower level
compared to the upper level and results in an occupation inversion situation where the energetically higher level has
higher occupation than the lower level. A second consequence of the small couplings $\Gamma_{1}$ is the sharp transition
of the occupation of the upper level from almost filled to almost empty. Once the upper level is almost empty the dot
system may gain energy by increasing the occupation of the lower level without having to pay Coulomb energy. This leads
to the nonmonotonic occupation of the lower level, known as charge oscillation. See Sindel et~al.\citep{sindel2005} for
a more detailed discussion. The results for the level occupation of the simple spinless model as shown in
\figref{fig:nrg-compare}, demonstrate excellent agreement between both NRG and DMRG calculations. The relative
difference of the ground-state energies obtained by NRG and MPS was on average $10^{-5}$.

\begin{figure}[t]
  \centering
  \includegraphics[width=\linewidth]{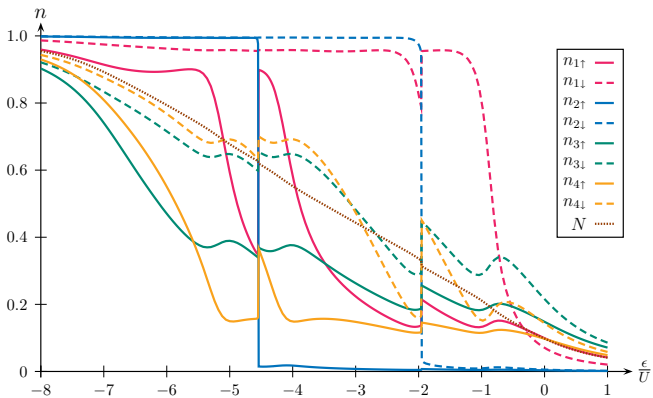}
  \caption{(Color online) Dot level occupation for a spinful four-level system. We parametrize the dot level energies as
    $\epsilon_{is} = \epsilon + \epsilon_{i} \pm B/2$ for $s = \shortuparrow,\shortdownarrow$, where $B$ represents the
    applied magnetic field with $B=0.2U$ and $\epsilon$ a gate voltage, with $\epsilon_{i} = (-0.1, -0.03, 0.07, 0.1)U$ 
    The coupling of the dot levels are chosen asymmetrically $\Gamma_{ir} = s_{i}\Gamma_{il}$ with $s_{i} = (1, -1, -1,
    1)$ and $\Gamma_{il} = (0.5, 0.02, 1, 0.7) 0.2U$.}
  \label{fig:4l-occ}
\end{figure}

We demonstrate the power of the MPS approach by considering a spinful four-level dot coupled asymmetrically to two
leads, a system sufficiently complex that its treatment by NRG is a highly challenging task. We therefore have no NRG
reference data for this system and present only DMRG results. For every dot level we calculate the
occupation $n_{is} = \expect{d_{is}^{\dagger}d_{is}^{\phantom{*}}}$ as a function of gate voltage, as shown in
\figref{fig:4l-occ}. This calculation is solely performed within the site optimization scheme. We kept the effective
dimensions for all $A$ matrices describing the leads the same compared to the two-level plot, only the LSS size at the
dot matrices was increased, thus demanding more computational time for the optimization at the dot.

For the four-level system we chose random values for the level couplings $\Gamma$ varying over two orders of
magnitude. Moreover, as the couplings have been chosen asymmetric, one \emph{cannot} simplify the model by decoupling
certain linear combinations of the leads, while keeping the remaining relevant degrees of freedom. The occupation of the
individual levels shows very rich behavior. By sweeping the gate potential similar to the spinless case above, we find
the sharpest transition for the second level ($n_{2\shortuparrow}, n_{2\shortdownarrow}$). The couplings of this level
are one magnitude smaller than all other couplings causing this sharp transition and associated with it charge
oscillations in all the other levels.

\section{Rotation to optimal basis of Wilson chains}
\label{sec:rotate-leads}

As described above the use of a star-shaped MPS works well for local quantities. However, one might ask the question
whether introducing such a geometry causes a loss of longer-ranged correlations between different chains.
To be able to assess this question we consider two sites in different chains ${\rm c}\neq {\rm c'}$, both at distance
$k$ from the dot. The mutual information\cite{NielsonChuang2000} $I_{\rho}^{\rm cc'}(k)$ contained between these two
sites is given by
\begin{equation}
  \label{eq:mutual-information}
  I_{\rho}^{\rm cc'}(k) = S_{\rho_{\rm red}^{\rm c^{\phantom{*}}}(k)} + S_{\rho_{\rm red}^{\rm c'}(k)} - S_{\rho_{\rm red}^{\rm cc'}(k)}
\end{equation}
with the entropy $S$
\begin{equation}
  \label{eq:entropy}
  S_{\rho} = -\tr\left(\rho\ln\rho\right).
\end{equation}
Here $\rho_{\rm red}^{\rm c}(k)$ is the reduced one-site density matrix obtained by tracing out the entire system except
for site $k$ in chain $c$. Likewise $\rho_{\rm red}^{\rm cc'}(k)$ is the reduced two-site density matrix, obtained
by tracing out all sites except two, situated at a distance $k$ from the dot in two different chains, $c$ and
$c'$. $I_{\rho}^{\rm cc'}(k)$ is a measure for how much information the sites contain about each other. As a
consequence, a decaying $I_{\rho}^{\rm cc'}(k)$ as a function of distance $k$ indicates that chains $c$ and $c'$
effectively decouple.
\begin{figure}[t]
  \centering
  \includegraphics[width=\linewidth]{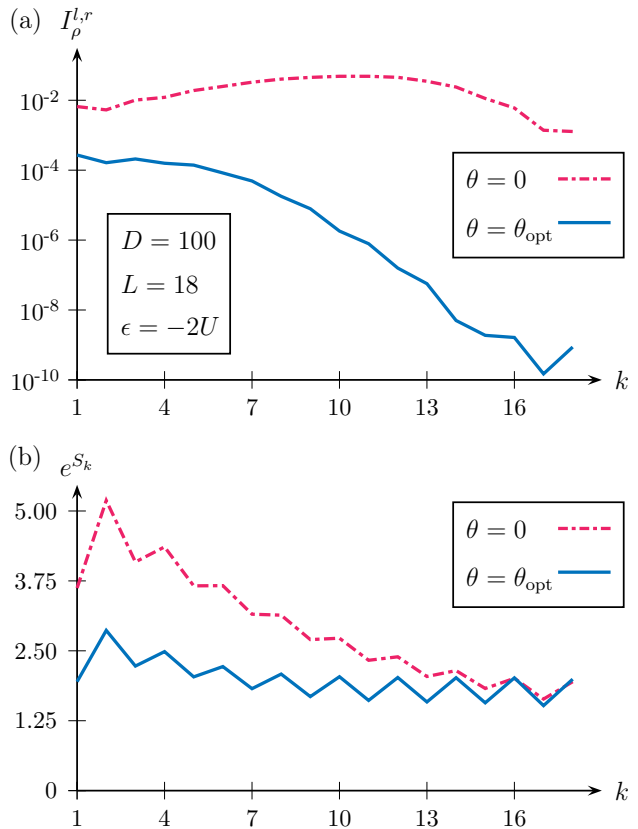}
  \caption{(Color online) (a) Mutual information $I_{\rho}^{l,r}$ between two sites situated in different leads but at equal
    distances $k$ from the dot, for a spinless four-level, two-lead model  with dot levels $\epsilon_{i}/U = (-0.1, -0.03,
    0.07, 0.1) + \epsilon$, $\epsilon=-2U$ fixed, couplings $\Gamma_{i r} = (0.3, -0.02, -1, 0.2)$ and $\Gamma_{i l} =
    (0.5, 0.08, 1, 0.7)$ and $\Lambda = 3$. The dashed line shows $I_{\rho}^{l,r}$ for the system with the leads in the
    original basis of Eq.~\eqref{eq:h-total}, whereas the solid line shows $I_{\rho}^{l,r}$ after the leads have been rotated by
    the (fixed, $k$ independent) optimal angle $\theta_{\rm opt}$ obtained from \figref{fig:rotate-leads-angle}(a).
    (b) Exponentiated bond entropy $e^{S_{k}}$ along the right chain of the system both prior (dashed line) and after
    (solid line) the rotation with $\theta_{\rm opt}$, indicating an effective reduction in the required matrix
    dimension $D_{k}$ close to the impurity for the rotated system by about $\frac{1}{2}$ for the same numerical
    accuracy.}
  \label{fig:rotate-mutinf}
\end{figure}

For simplicity and to make a comparison with NRG feasible, we restrict ourselves to the spinless case, e.g.,\ we
only look at the spin-up part of the original four-level system, however with different couplings compared to the
parameters used for \figref{fig:4l-occ}. As NRG treats both the left and right lead in a combined single chain we can,
nevertheless, study the effect of ``unfolding'' the two parts of the NRG chain.

If we calculate $I_{\rho}^{l,r}$ for this spinless two-lead Hamiltonian as it stands, the correlations
between two sites on opposite sides of the dot but at equal distance from it are found to decay only very weekly with
$k$ [\figref{fig:rotate-mutinf}(a), dot-dashed line]. This illustrates, on the one hand, that our MPS ansatz does
successfully capture correlations between sites representing comparable energy scales, in spite of the fact that in the
star geometry they lie ``far'' from each other (namely on different chains). On the other hand, it also raises the
question whether one can choose a (numerically) better suited basis for the leads that effectively does decouple
different chains far from the dot. Since in that case the correlations would intrinsically decay with distance from the
dot, less numerical resources would be required to capture all correlations accurately.

Indeed, we shall show that it is possible to choose such an optimal basis by making a suitably chosen unitary
transformation which rotates the lead degrees of freedom into each other in an ``optimal'' way to be described
below. When the leads are first rotated by a certain optimal angle of rotation $\theta_{\rm opt}$ (defined precisely
below) and $I_{\rho}^{l,r}$ is calculated in this rotated basis, then $I_{\rho}^{l,r}$ is found to decay rapidly with
$k$, see solid line in \figref{fig:rotate-mutinf}(a).

\begin{figure}[t]
  \centering
  \includegraphics[width=\linewidth]{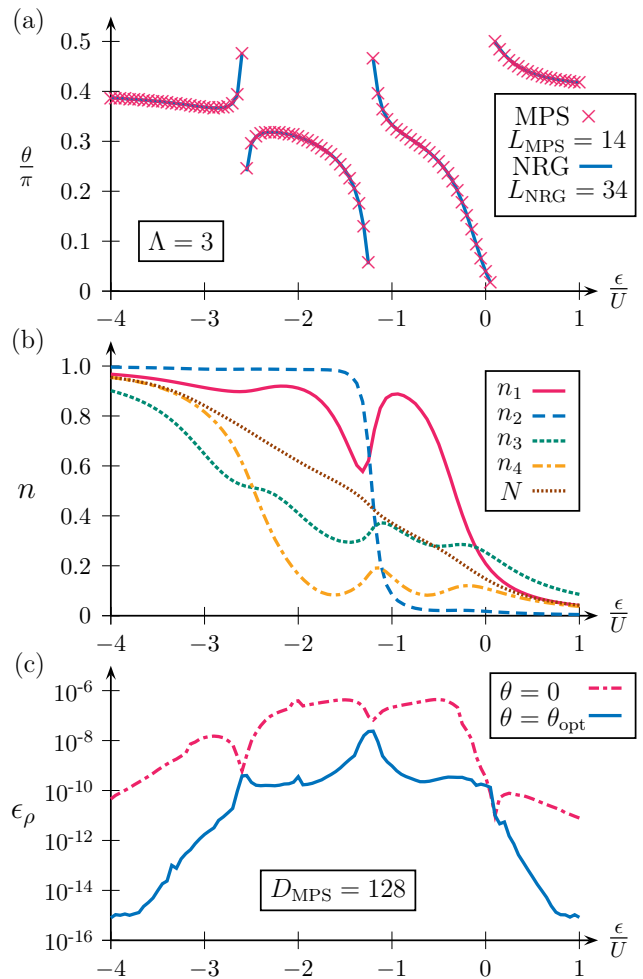}
  \caption{(Color online) Optimal basis for the leads of a spinless four-level, two-lead system (same parameters as for
    \figref{fig:rotate-mutinf}, but with varying $\epsilon$). (a) Optimal angle of rotation $\theta_{\rm opt}$ for the
    leads obtained by diagonalizing $\rho_{\rm red}^{l,r}(k=L)$ for the DMRG calculation (red symbols) in comparison
    with angle that diagonalizes the scattering matrix calculated with NRG (blue line). $\theta_{\rm opt}$ is defined
    $\mod\pi/2$. (b) Dot level occupation. $N = \frac{1}{4}\sum_{i=1}^{4}n_{i}$ is the rescaled total dot
    occupation. Rapid changes in the angle $\theta_{\rm opt}$ coincide with rapid shifting of dot-level occupations. (c)
    Truncation error (accumulated discarded density-matrix eigenvalues) of the DMRG calculation considering two
    neighboring sites at a time for a rotated and non-rotated system. We typically used 20 sweeps for the DMRG
    calculations. The truncation error is significantly reduced for the rotated system except for the points where
    $\theta_{\rm opt}$ actually shows rather rapid transitions through $\theta_{\rm opt}=0$ itself. At these points the
    leads are already decoupled from the outset.}
  \label{fig:rotate-leads-angle}
\end{figure}
We begin with the observation that the labeling of the unfolded chains with $\alpha = l, r$ is arbitrary. We can
choose any linear combination of $l$ and $r$ as new basis, e.g.,\ for symmetric couplings to the dot it is well
known that with the symmetric and antisymmetric combination only the symmetric lead couples to the dot while the
antisymmetric lead is completely decoupled. To be specific, we can introduce a unitary transformation acting on the
original lead states specified in the Hamiltonian
\begin{equation}
  \label{eq:rotate-leads}
  \tilde{f}_{\beta n\sigma} = U_{\beta\alpha} f_{\alpha n\sigma}
\end{equation}
independent of the site $n$ and spin $\sigma$, acting only on the lead index $\alpha$. For systems with time-reversal
symmetry, the unitary matrix is always chosen real. So in our case, for $N_{\rm l}=2$ $U = U(\theta)$ is a real
two-dimensional matrix and can be thought of as a planar rotation parametrized by a single angle $\theta$. The optimal
basis for DMRG treatment would have minimal correlations between the rotated chains. The angle of rotation $\theta$ can
be restricted to $\theta\in [0,\pi/2]$ as we choose to ignore the particular order and relative sign of the new basis
vectors. In order to find the optimal angle it is sufficient to look at the reduced two-site density matrix $\rho_{\rm
  red}^{l,r}(k)$. As the Hamiltonian \eqref{eq:h-total} preserves particle number, this density matrix is a $4\times 4$
matrix in block form: a $1\times 1$ block for both the zero-particle and two-particle sectors and a $2\times 2$ block
for the one-particle sector.

Finite off-diagonal elements of this $2\times 2$ block show that both sites are effectively correlated with each
other. However, by diagonalizing this block of $\rho_{\rm red}^{l,r}(k)$ via a real unitary matrix $U$ we immediately
obtain a rotated lead basis according to Eq.~\eqref{eq:rotate-leads}. So the angle of rotation $\theta_{\rm opt}$ can be
found by diagonalizing $\rho_{\rm red}^{l,r}(k)$. It is most desirable to decouple the far ends of the chains best, so
we choose $\theta = \theta(k=L)$, where $U[\theta(k=L)]$ diagonalizes $\rho_{\rm red}^{l,r}(k=L)$.

By applying the transformation $U(\theta)$ to the Hamiltonian \eqref{eq:h-total} only the tunneling elements to and
from the dot levels are changed
\begin{equation}
  \label{eq:rotate-ham-change}
  \tilde{V}_{\beta i\sigma} = U(\theta_{\rm opt})_{\beta\alpha}V_{\alpha i\sigma}.
\end{equation}
This way, we have obtained a new lead basis for our Hamiltonian that is better suited for the DMRG calculations, as
long ranging correlations are suppressed in this basis. As we benefit already from a rotation in the leads even if the
angle is only close (but not equal) to the optimal choice $\theta_{\rm opt}$, it is feasible to start with a small
system (of only, say, 14 sites per Wilson chain) in order to obtain an approximate value for $\theta_{\rm opt}$; the
latter can then be used to rotate the leads of a bigger system, from which a better determination of the optimal angle
can be extracted.

In \figref{fig:rotate-leads-angle} we show the optimal angle of rotation $\theta_{\rm opt}$ for a spinless four-level
system. We compare with NRG calculations where we diagonalize the $T$-matrix
\begin{equation}
  \label{eq:nrg-t-matrix}
  T_{\alpha\beta} = \lim_{\omega\rightarrow 0^{+}}\left(\hat{V}^{\dagger}\hat{G}^{\rm dot}(\omega)\hat{V}\right)_{\alpha\beta},
\end{equation}
where $\hat{G}^{\rm dot}$ is the local retarded Green's-function matrix calculated by standard NRG
techniques\cite{WeichselbaumDelft2007} and $\hat{V}$ is the tunneling matrix from the Hamiltonian. The angle extracted
from the diagonalization of the $T$ matrix [i.e.,\ from requiring that $U(\theta) T U^\dagger (\theta)$ be diagonal] is
shown as a solid line in panel (a). Remarkably, this line agrees quantitatively with the $\theta_{\rm opt}$ values found
by DMRG. This shows that the angle of rotation that minimizes correlations between the two rotated leads has a clear
physical interpretation: it also diagonalizes the scattering matrix, a result that is intuitively very reasonable.
We note, though, that this fact cannot be used to determine $\theta_{\rm opt}$ \emph{before} doing the DMRG calculation,
as with the knowledge of the scattering matrix we would have already solved the system. Nevertheless, shorter systems
can already give a clean indication of the angle that decouples the chains.

In \figref{fig:rotate-mutinf} we demonstrate that by rotating the leads to the new optimal basis as suggested above it
is possible, indeed, to ensure that lead degrees of freedom on different (rotated) Wilson chains become effectively
decoupled from each other further out on the chains. Also the bond entropy $S_{k}$ is reduced. If the leads are rotated
into the optimal basis the mutual information drops quickly along the chains [see \figref{fig:rotate-leads-angle}(a)],
and the truncation error is significantly smaller [see \figref{fig:rotate-leads-angle}(c)], thus making numerical
treatment less demanding. Note that rapid changes in the angle $\theta_{\rm opt}$ coincide with rapid shifting of
dot-level occupations [see \figref{fig:rotate-leads-angle}(b)].

\section{Summary}
\label{sec:summary}

Using the DMRG approach gives us the possibility to choose a more flexible MPS geometry compared to
NRG. While in NRG one is bound to a simultaneous treatment of a single combined Wilson chain due to the requirement of
energy scale separation, this restriction can be lifted in a DMRG treatment. In our case of a two-lead Anderson model we
modeled each spin and each lead by a Wilson chain on its own treated separately from each other. Thus we achieved a
significant reduction in both the dimension of the LSS and the dimension $D$ of the ISS and OSS at each site. The
Hilbert space of one site in the single chain geometry is equivalent to the direct product of the Hilbert spaces of the
$4 = 2 N_{\rm l}$ corresponding sites of each chain. So in order to map a star geometry description into an equivalent
single chain description, in the sense that the effective Hilbert spaces at every site have the same dimension, the
dimension $D'$ of the single chain $A$ matrices would scale exponentially with the number of leads, $D' \simeq D^{2
  N_{\rm l}}$, as a consequence of the tensor product of the $2N_{\rm l}$ smaller star geometry $A$ matrices. Thus, adopting the
star geometry reduces the numerical costs for treating the leads by $D \simeq {D'}^{1/2N_{\rm l}}$. Although this strategy has
the consequence that the cost of treating the dot site increases significantly [see Eq.~\eqref{eq:cost-mult}], for all
cases studied in this paper the latter effect is far outweighed by the decrease in costs for treating the leads. Indeed
we found that dimensions $D \le 36$ suffice for getting an accurate description of the system (note that when translated
into a single chain this would result in a huge effective dimension of $D' \simeq D^4 = 1.7 \times 10^6$).

Due to the fact that the Anderson Hamiltonian under consideration features only a density-density interaction term at
the dot between electrons with different spin, the dot matrices can be conveniently split into two sets, one for each
spin, yielding another gain in efficiency. As it turns out, the dimension $D_{v}$ connecting to two sets of dot
matrices can be chosen significantly smaller than $D$ (cf. \figref{fig:bond-dim}).
In addition an optimal basis (in terms of numerical efficiency) for representing the leads has been determined, which
minimizes correlations between different Wilson chains and in which, it turns out, the scattering matrix becomes
diagonal. Moreover, the DMRG sweeping procedure allows the dimensions $D$ of the MPS matrices in the system to be
adjusted very flexibly. Indeed, in our case it was possible to reduce the matrix dimensions along the chain away from
the dot considerably. The combination of all these resource-saving features makes it feasible to calculate the
\emph{ground-state} properties of generic complex quantum impurity models using only relatively moderate numerical
resources.

The calculation of dynamical quantities like local spectral functions is, in principle, also possible for the star
geometry, for example, by suitably modifying the approach of Ref.~\onlinecite{verstraete:vmps} to the present
geometry. However, we expect that the increased computational costs of DMRG relative to NRG for calculating dynamical
quantities would in this case likely offset the advantages of the star geometry.

In closing, we would like to make the following comment: while we expect that a rotation to an optimal basis as
described above should be applicable to a large class of impurity models, there may be cases where it does not work. In
particular, we suspect that this might be the case for some models showing non-Fermi-liquid behavior, such as the
two-channel spin-$\frac{1}{2}$ Kondo model, where overscreening of the impurity is likely to lead to strong mutual correlations
between all Wilson chains. A quantitative analysis of this problem using the present star geometry approach is beyond
the scope of the present investigation but would be an interesting subject for future study.

\begin{acknowledgments}
We gratefully acknowledge fruitful discussions with Frank Verstraete,
Ulrich Schollwöck, Theresa Hecht, Hamed Saberi, and Ian McCulloch. This work was
supported by DFG (SFB 631, De-730/3-2, SFB-TR12, SPP 1285, De-730/4-1).
Financial support of the German Excellence Initiative via the ``Nanosystems
Initiative Munich (NIM)'' is gratefully acknowledged.
\end{acknowledgments}


\end{document}